\documentclass[floatfix,superscriptaddress,a4paper,
               showpacs,showkeys,nofootinbib,twocolumn]{revtex4}
\usepackage{epsfig}
\usepackage{latexsym}
\usepackage{xspace}
\usepackage{hyperref}
\usepackage[latin2]{inputenc}
\usepackage{indentfirst}
\usepackage{float}

\usepackage{xcolor}

\usepackage{amsmath}
\usepackage{amssymb}
\usepackage[english]{babel}
\usepackage{url}

\graphicspath{{./},{./graphics/}}

\begin{document}

\title{Event-by-event fluctuations in p+p and central A+A collisions\\ within relativistic transport models}
\author{Anton Motornenko}
\affiliation{Frankfurt Institute for Advanced Studies, Giersch Science Center, Frankfurt am Main, Germany}
\affiliation{Institut f\"ur Theoretische Physik,
Johann Wolfgang Goethe Universit\"{a}t, Frankfurt am Main, Germany}

\author{Katarzyna Grebieszkow}
\affiliation{Faculty of Physics, Warsaw University of Technology, Warsaw, Poland}

\author{Elena Bratkovskaya}
\affiliation{Institut f\"ur Theoretische Physik,
Johann Wolfgang Goethe Universit\"{a}t, Frankfurt am Main, Germany}
\affiliation{GSI Helmholtzzentrum f\"{u}r Schwerionenforschung GmbH, Darmstadt, Germany}

\author{Mark I. Gorenstein}
\affiliation{Frankfurt Institute for Advanced Studies, Giersch Science Center, Frankfurt am Main, Germany}
 \affiliation{Bogolyubov Institute for Theoretical Physics, Kiev, Ukraine}

\author{Marcus Bleicher}
\affiliation{Frankfurt Institute for Advanced Studies, Giersch Science Center, Frankfurt am Main, Germany}
\affiliation{Institut f\"ur Theoretische Physik,
Johann Wolfgang Goethe Universit\"{a}t, Frankfurt am Main, Germany}
\affiliation{GSI Helmholtzzentrum f\"{u}r Schwerionenforschung GmbH, Darmstadt, Germany}

\author{Klaus Werner}
\affiliation{Laboratoire SUBATECH, University of Nantes - IN2P3/CNRS - Ecole desMines,
Nantes, France}

\begin{abstract}
Event-by-event multiplicity fluctuations in
nucleus-nucleus collisions are studied within the relativistic transport models:
EPOS, PHSD, and UrQMD. As measures of particle number fluctuations we consider
the scaled variances $\omega[X]$ for positive, negative, and all charged hadrons,
and the strongly intensive
quantities
$\Delta[K^+,\pi^+],$ $\Sigma[K^+,\pi^+]$ for $K^+$ and $\pi^+$ yields.
At the SPS energy range the fluctuation measures
are calculated for proton-proton,  Ar+Sc, and Pb+Pb collisions.
Comparison with recent NA61/SHINE and older NA49 measurements of the multiplicity
fluctuations
is done. A validity of the model of independent sources, a role of the experimental acceptance, and
the centrality selection procedure are studied.

\end{abstract}
\date{\today}

\maketitle

\section{Introduction}
The key questions of current research in high energy nuclear physics are properties of strongly-interacting matter and its transition to quark-gluon plasma (QGP) state. Fluctuations are expected to play an important role in study of the phase structure of QCD matter formed in relativistic heavy ion collisions \cite{Heiselberg:2000fk, Jeon:2003gk,Asakawa:2015ybt}. In the vicinity of the critical point fluctuations of particle multiplicity are significantly enlarged. This gives an opportunity to relate experimental measures to critical behavior of system created in nucleus-nucleus collision. 
The study of multiplicity fluctuations is a complicated experimental procedure because of the finite detector acceptance and limited particle identification. 
It is important also to eliminate event-by-event volume fluctuations that are a trivial source of multiplicity fluctuations and are not related to transition between phases.

The relativistic transport models such as
EPOS~\cite{Werner:2005jf, Werner:2010aa}, PHSD \cite{Cassing:2008sv,Cassing:2009vt},
and UrQMD \cite{Bass:1998ca, Bleicher:1999xi} allow to perform theoretical study of
microscopical properties of systems that are created in nucleus-nucleus (A+A) collisions.
This detailed description of A+A collisions allows to study the particle number fluctuations under different conditions, see, e.g.,
\cite{Konchakovski:2005hq, Konchakovski:2006aq, Gorenstein:2008et,Konchakovski:2010fh, Begun:2012wq, Vovchenko:2014ssa, Steinheimer:2016cir}.
Recent data from the NA61/SHINE Collaboration~\cite{Seryakov:2017sss} show a
surprisingly different behavior of scaled variances $\omega$ for
proton-proton (p+p) and A+A collisions. It has been found that multiplicity fluctuations
are essentially larger in p+p inelastic collisions than in most central A+A ones.
This result is in a contradiction with the wounded nucleon model.
An aim of the present work is to investigate a dependence
of the multiplicity fluctuations on the system volume and collision energy within the
microscopical transport models. We also compare the results of the transport
models with the existing data.

The paper is organized as follows. In Sec. II the measures of particle number fluctuations are
introduced. In Sec. III a short reminding of the basic concepts of three popular transport models
is presented. The main results are considered in Sec. IV, and Sec. V summarizes the paper.

\section{Event-by-event fluctuations in transport models}
In the present study the simulations of A+A collisions were performed
within the transport models listed in Sec. I. The calculations are done
for systems of different size, i.e., p+p, Ar+Sc, and Pb+Pb are studied.
The collision energies are taken in a range of the CERN SPS accelerator
which corresponds to the center of mass energy of nucleon pair
$\sqrt{s_{NN}}=5.1-17.3$~GeV.
The specific energy values in the present study are selected to coincide
with those in the NA61/SHINE experiment (see, e.g.,  Refs.~\cite{Antoniou:2006mh, Abgrall:2014xwa}).
Note that in the aforementioned experiment the centrality selection
is a rather complicated procedure
defined by the number of projectile spectators \cite{Seryakov:2017sss, Seryakov-private}.
To model this centrality selection in Ar+Sc collisions
we use the recently introduced experimental acceptance maps of the forward energy spectrometer.

The following measures of particle number fluctuations  are studied in the present work:
\begin{gather}
\omega\left[X\right] = \dfrac{\left<X^2\right> - \left<X\right>^2}{\left<X\right>}\, ,\label{omega}\\
\Delta[A,B]
 ~=~ \frac{1}{C_{\Delta}} \Big[ \langle B\rangle\,
      \omega[A] ~-~\langle A\rangle\, \omega[B] \Big]~, \label{Delta}\\
\begin{split}
\Sigma[A,B]
 ~=~ \frac{1}{C_{\Sigma}}\Big[
      \langle B\rangle\,\omega[A] ~+~\langle A\rangle\,\omega[B] ~-\\
      -~2\left(
      \langle AB \rangle -\langle A\rangle\langle
      B\rangle\right)\Big]\, ,\label{Sigma} \\
\end{split}\\
      C_\Delta = \langle A \rangle\ - \langle B \rangle\, , ~~
      C_\Sigma = \langle A \rangle\ + \langle B \rangle\, ,\nonumber
\end{gather}
where $X,$ $A$, and $B$ denote the particle yields, and
\begin{equation}
\langle X \rangle~=~\frac{1}{N_{\rm ev}}\sum_{i=1}^{N_{\rm ev}} X_i
\end{equation}
corresponds to the event-by-event averaging over a sample of $N_{\rm ev}$ events.
The  fluctuation measures (\ref{omega}-\ref{Sigma}) are intensive quantities, i.e.,  they do
not depend on the system volume.
However, only $\Delta$ (\ref{Delta}) and $\Sigma$ (\ref{Sigma}),
are called as strongly intensive quantities \cite{Gorenstein:2011vq},
because they are not sensitive to the fluctuations of the system volume. The system volume event-by-event
fluctuations in A+A  collisions are usually a result of the varying impact parameter from collision to collision.
Note that even at the fixed value of the impact parameter $b=0$ the number of nucleon participants
still fluctuates event-by-event, and these fluctuations influence the scaled variances $\omega[X]$. 

The scaled variances $\omega$ of distribution particle numbers are used in a large number of both theoretical
and experimental papers, see, e.g., the reviews \cite{Jeon:2003gk,Konchakovski:2006aq}. The strongly intensive measures were introduced in \cite{Gorenstein:2011vq}.
They are actively used now by the NA49 and NA61/SHINE collaborations for the analysis of event-by-event fluctuations
in p+p and nucleus-nucleus collisions in the SPS energy region.

In the present work the following goals are pursued:
\begin{itemize}
\item Study of particle number fluctuations versus collision energy and system size.
Search for any non-monotonous behavior.
\item Comparison of different transport models.
\item Investigation of the role of different acceptance criteria.
\item Comparison of the transport model results for $\omega[N_-]$
with the recent data of the NA61/SHINE Collaboration in p+p, Ar+Sc, and
the NA49 data in Pb+Pb collisions.
\end{itemize}

There are two well known assumptions that connect
the scaled variances obtained
in  different physical scenarios.
The first one (see, e.g., \cite{Begun:2004zb}) presents the scaled variance
$\omega_{\rm acc}[X]$ for experimentally accepted particles (e.g.,
in the finite
kinematical region of a momentum space) in terms of $\omega[X]$ which describes
the event-by-event fluctuations in the case when all particles are accepted:
\begin{equation}
\omega_{\rm acc}[X]=1-q+q\,\omega[X],~~~~0<q=\frac{\left<X_{\rm acc}\right>}{\left<X\right>}<1\, ,
\label{eq:omega-acc}
\end{equation}
where $q$ is the acceptance parameter equal to the ratio of the mean values
of the accepted number of particles to the total one.
Equation (\ref{eq:omega-acc}) ignores possible correlations between
accepted particles. It predicts
that at $q\rightarrow 0$ the scaled variance
$\omega_{\rm acc}$ goes {\it monotonously}  to 1,
the value that corresponds to the
Poisson distribution. Thus, $\omega_{\rm acc}[X]$ increases linearly with $q$,
if $\omega[X]>1$, or decreases linearly with $q$, if $\omega[X]<1$.

The second assumption often used in A+A collisions
is the model of independent sources, e.g., the so-called wounded nucleon model~\cite{Bialas:1976ed}.
This model connects
the scaled variance $\omega[X]$ for particles produced
in A+A collisions with the scaled variance $\omega_{\rm NN}[X]$ and
average multiplicity $\langle X\rangle_{\rm NN}$ taken from  nucleon-nucleon (NN)
collisions at
the same collision energy $\sqrt{s_{NN}}$:
\begin{equation}\label{part}
\omega[X] ~=~ \omega_{\rm NN}[X]~ + ~\omega_{\rm p}[N_{\rm part}]\,\frac{1}{2}\,\langle X\rangle_{\rm NN}\, ,
\end{equation}
where $\omega_{\rm p}[N_{\rm part}]$ represents the scaled variance of distribution of nucleon participants number.

As it was already pointed out in Sec. I, the recent data of NA61/SHINE Collaboration for $\omega[N_-]$
in p+p and central Ar+Sc collisions are in a contradiction with Eq.\ref{part}. An experimental
verification of Eq.\ref{eq:omega-acc} looks still problematic. In what follows we will check
whether Eq.\ref{eq:omega-acc} and Eq.\ref{part} are valid within transport model calculations.

\section{Transport Models}
\subsection{EPOS}

In the Energy Parton Off-shell Spliting (EPOS) transport model,
nucleus-nucleus scattering - even proton-proton - amounts
to many elementary collisions happening in parallel. Such an elementary
scattering corresponds to the exchange of a ``parton ladder''.

A parton ladder represents parton evolution from the projectile and
the target side towards the center (small $x$). The evolution is
governed by an evolution equation, in the simplest case according
to DGLAP. In the following we will refer to these partons as {}``ladder
partons'', to be distinguished from {}``spectator partons'' to
be discussed later. It has been realized a long time ago that such
a parton ladder may be considered as a quasi-longitudinal color field,
a so-called ``flux tube''~\cite{Werner:2010aa}, conveniently treated as a
relativistic string. The intermediate gluons are treated as kink singularities
in the language of relativistic strings, providing a transversely
moving portion of the object. This flux tube decays via the production
of quark-antiquark pairs, creating in this way fragments -- which
are identified as hadrons.

The technical details of the
consistent quantum mechanical treatment of the multiple scattering
with the energy sharing between the
parallel scatterings can be found in~\cite{Drescher:2000ha}.
 Hard scale independent correction factors are
added to the bare amplitude of the Pomeron to control the rise of the
cross-section at high energy and the multiplicity in HI collisions. The treatment of these
nonlinear effects at high energy is explained in \cite{Werner:2005jf}.
In the present paper we use the newest version of the EPOS model, the so called
EPOS-LHC in the CRMC mode.

\subsection{PHSD}
The Parton-Hadron-String Dynamics (PHSD) transport approach~\cite{Cassing:2008sv,Cassing:2009vt,Bratkovskaya:2011wp,Cassing:2008nn} is a microscopic covariant dynamical model for strongly interacting systems formulated on the basis of Kadanoff-Baym equations
\cite{Kadanoff1,Cassing:2008nn}.
The approach consistently describes a full evolution of the relativistic heavy-ion collision from the initial hard scatterings and string formation through the dynamical deconfinement phase transition to the strongly-interacting quark-gluon plasma (sQGP) as well as hadronization and the subsequent interactions in the expanding hadronic phase as in the Hadron-String-Dynamics (HSD) transport approach \cite{Cassing:1999es}.
The transport theoretical description of quarks and gluons in the PHSD is based on the Dynamical Quasi-Particle Model (DQPM) for partons that is constructed to reproduce lattice QCD for QGP thermodynamics~\cite{Cassing:2008nn,Berrehrah:2016vzw} via effective propagators for quarks and gluons.
The PHSD differs from conventional Boltzmann approaches in following aspects:\\
i) it incorporates dynamical quasi-particles due to the finite width of the spectral functions;\\
ii) it involves scalar mean-fields that substantially drive the collective flow in the partonic phase;\\
iii) it is based on a realistic equation of state from lattice QCD and thus describes the speed of sound $c_s(T)$ reliably;\\
iv) the hadronization is described by the fusion of off-shell partons to off-shell hadronic states and does not violate the second law of thermodynamics;\\
v) the effective partonic cross sections are self-consistently determined within the DQPM and probed by transport coefficients in thermodynamic equilibrium (shear- and bulk viscosity, electric conductivity, magnetic susceptibility etc.
\cite{Ozvenchuk:2012kh,Linnyk:2015rco}).

At an early stage of relativistic heavy-ion collisions color-neutral strings (described by the LUND model~\cite{Andersson:1992iq}) are produced in highly energetic scatterings of nucleons from the impinging nuclei. These strings are dissolved into 'pre-hadrons'.
If the local energy density is larger than the critical value for the phase transition, which is taken to be $\sim$ 0.5 ${\rm GeV/ fm^3}$, the pre-hadrons melt into (colored) effective quarks and antiquarks in their self-generated repulsive mean-field as defined by the DQPM~\cite{Cassing:2008nn,Berrehrah:2016vzw}. In the DQPM the quarks, antiquarks, and gluons are dressed quasi-particles and have temperature-dependent effective masses and widths which have been fitted to lattice thermal quantities such as energy density, pressure and entropy density.

The transition from the partonic to hadronic degrees-of-freedom
is described by covariant transition rates for the fusion of quark-antiquark pairs to mesonic resonances or three quarks (antiquarks) to baryonic states.
In the hadronic phase PHSD is equivalent to the
HSD model \cite{Cassing:1999es, Cassing:2000bj}.
The PHSD approach has been applied to p+p, p+A, and A+A collisions from lower SPS to LHC energies and been successful in describing of experimental data including single-particle spectra, collective flow as well as electromagnetic probes~
\cite{Cassing:2009vt,Bratkovskaya:2011wp,Konchakovski:2014fya,Konchakovski:2011qa,Linnyk:2015rco,Palmese:2016rtq}.

\subsection{UrQMD}
The Ultra relativistic Quantum Molecular Dynamics (UrQMD) model \cite{Bass:1998ca, Bleicher:1999xi} is a microscopic transport theory based on the covariant propagation of all hadrons on classical trajectories in combination with stochastic binary scatterings, color string formation and resonance decay. It represents a Monte Carlo solution of a large set of coupled partial integro-differential equations for the time evolution of the various phase space densities $f_i(x,p)$ of particle species $i=N,\Delta,\Lambda,$ etc.

The exchange of electric and baryonic charge, strangeness and four-momentum in the  $t$-channel is considered for baryon-baryon (BB) collisions at low energies, while meson-baryon (MB) and meson-meson (MM) interactions are treated via the formation and decay of resonances, i.e. the $s$-channel reactions. $t$-channel reactions for MB and MM collisions are taken into account from $\sqrt s> 3$~GeV on increasing to the only MB, MM interaction type above $\sqrt s = 6$~GeV.

This framework allows bridging with one concise model the entire available range of energies from the SIS energy region ($\sqrt{s} \approx 2\,$GeV) to the RHIC energy ($\sqrt{s} = 200\,$GeV). At the highest energies, a huge number of different particle species can be produced. The model should allow for subsequent rescatterings. The collision term in the UrQMD model includes more than fifty baryon species and five meson nonets (45 mesons). In addition, their antiparticles have been implemented using charge-conjugation to assure full baryon-antibaryon symmetry.

All particles can be produced in hadron-hadron collisions and can interact further with each other. The different decay channels all nucleon-, $\Delta$- and hyperon-resonances up to 2.25$\,$GeV/$c^2$ mass as well as the meson (e.g. K$^*$) decays etc. are implemented.  At higher energies the advantage of the hadron universality is taken and a string model for the decay of intermediate states is used.

\section{Results}

The simulations of A+A collisions are performed
within the certain range of energies and system sizes:
p+p collisions are studied at energies $\sqrt{s_{NN}}=6.27,~7.75,~8.77,~12.33,~17.28$ GeV,
while for Ar+Sc and Pb+Pb collisions energies of the colliding system are
$\sqrt{s_{NN}}=6.12,~7.62,~8.77,~11.94,~16.84$ GeV.
These are the energies that are available in the NA61/SHINE experiment.

Different acceptance cuts are applied for the transport model simulations.
The results are presented in three different acceptance regions:
\begin{itemize}
\item full rapidity ($4\pi$) -- all particles are accepted in each event;
\item $|y|<1$ acceptance -- only particles with rapidity $-1<y<1$ in the
center of mass frame are accepted;
\item the NA61/SHINE acceptance  -- $\pi^\pm$, $K^\pm$, $p$, and $\bar{p}$ are accepted
in the forward rapidity region $0<y<y_{\rm beam}$
and transverse momenta $p_T<1.5$ GeV/c.
Additionally, the so-called NA61/SHINE acceptance map \cite{NA61-acceptance}
is then applied in order to take into account the NA61/SHINE acceptance in the azimuthal angle.
\end{itemize}

The number of events was always taken large enough, so that the statistical errors of our transport model simulations
are smaller than the size of markers on the plots. However, an estimation of the systematic errors in the transport
models is rather problematic and they are not shown in the presented plots. This is because a lot of uncertainties,
both experimental and theoretical, within each of the considered models,  e.g. the treatment of the jet production in hard collisions
are quite differ in the transport approaches which also contributes to the systematic errors.

\subsection{Scaled variance for particle number fluctuations}

\begin{figure*}[t]
\centering
    \includegraphics[width=.7\textwidth]{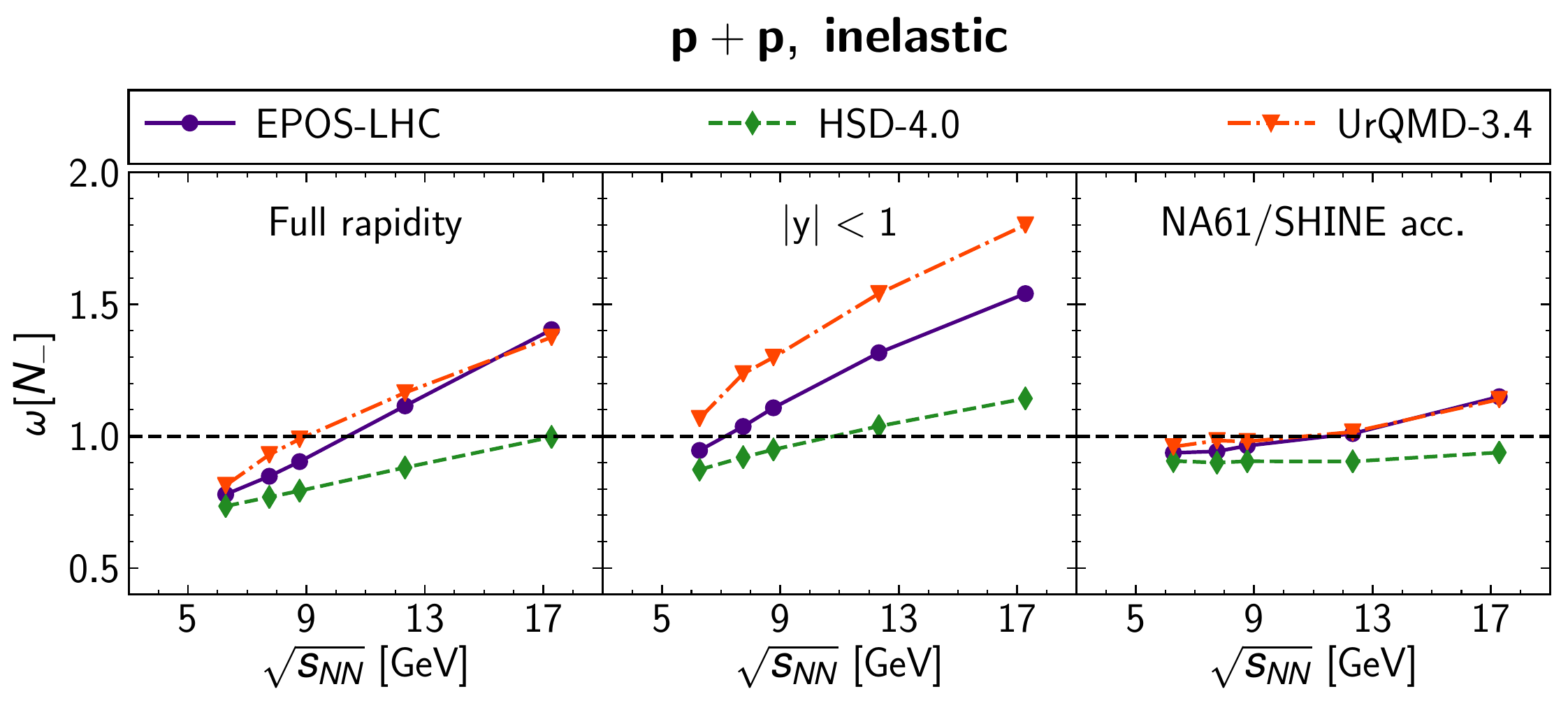}
    \includegraphics[width=.7\textwidth]{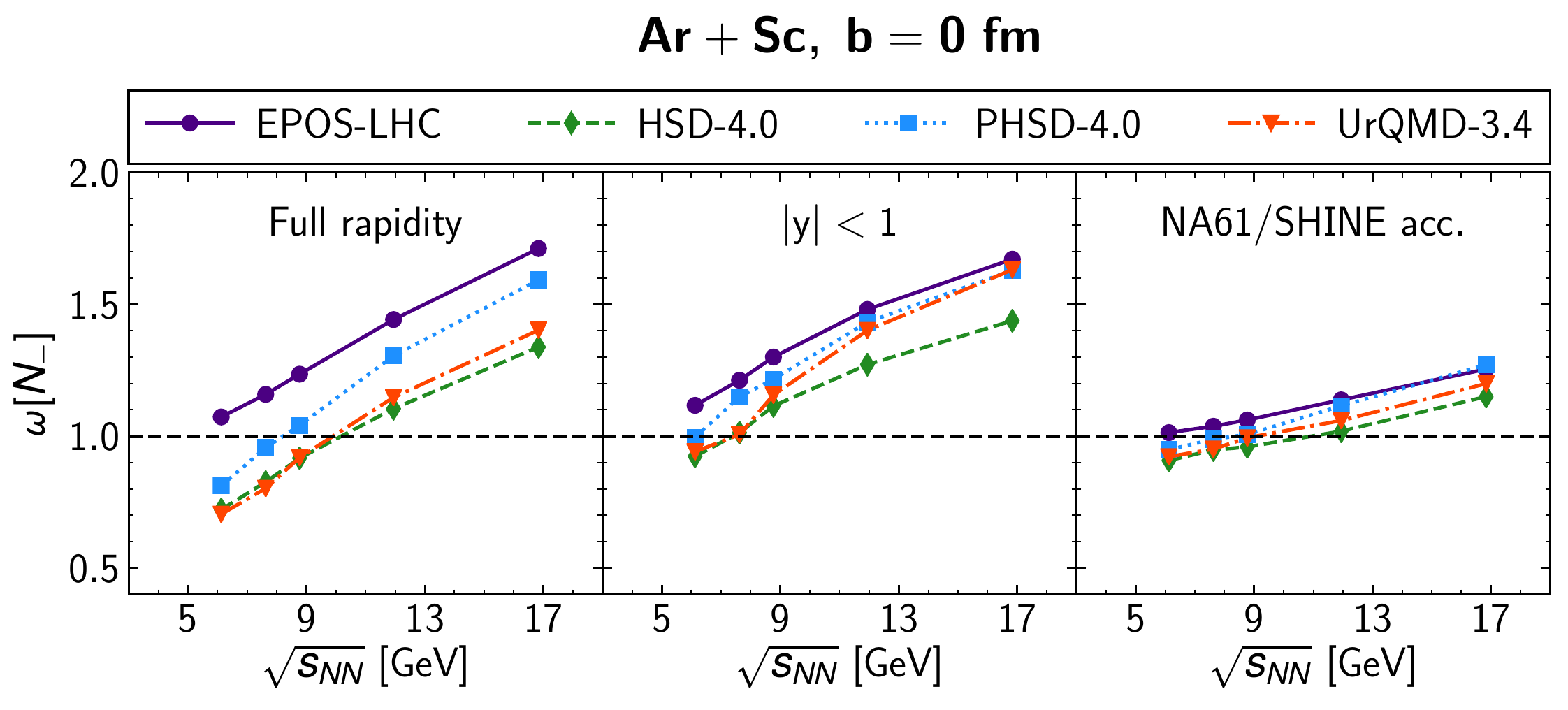}
    \includegraphics[width=.7\textwidth]{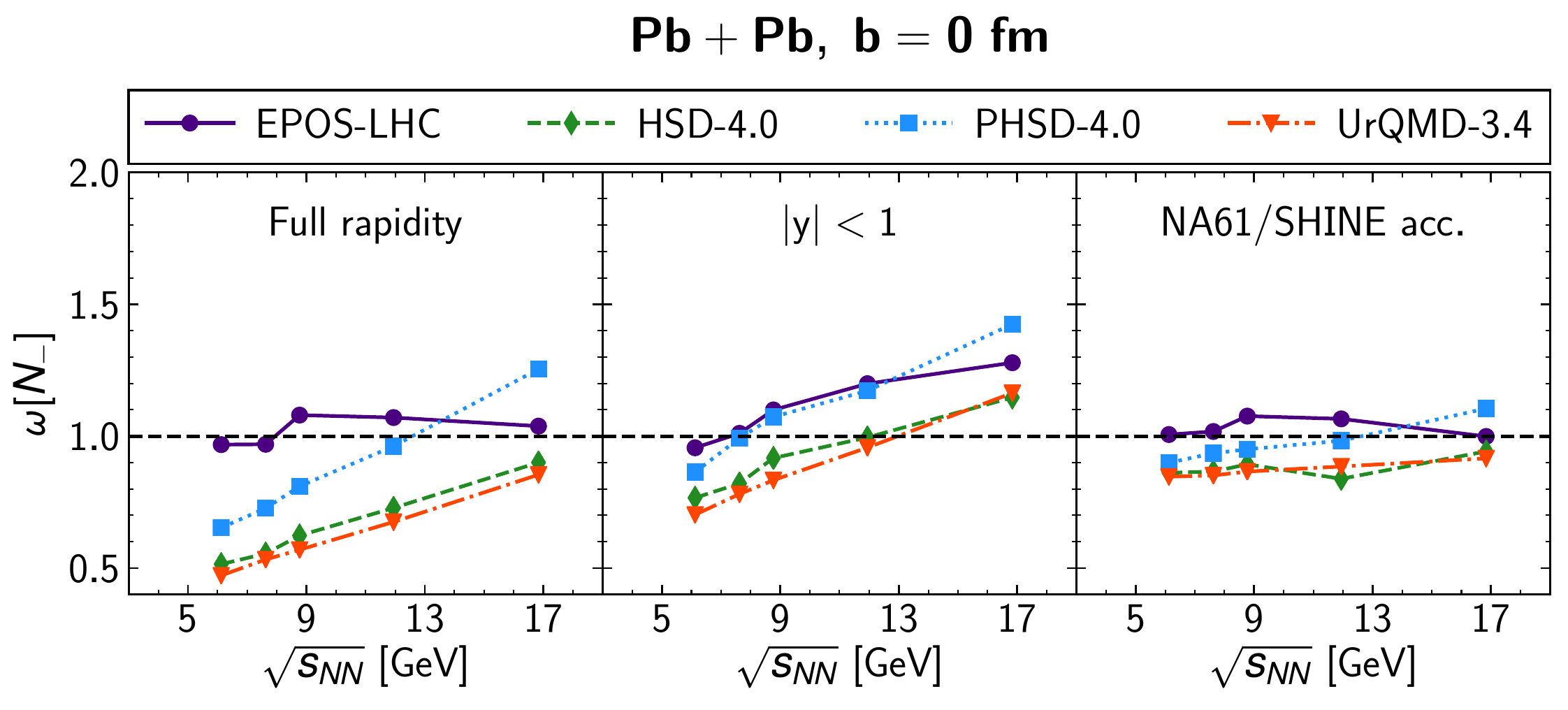}
    \caption{Comparison of EPOS (purple circles), HSD (green diamonds), PHSD (blue squares), and UrQMD (red triangles) results for scaled variance $\omega[N_{-}]$ of distribution of negatively charged particles number $N_{-}$ in inelastic p+p collisions ({\bf top}), $b=0$ fm Ar+Sc collisions ({\bf middle}), and $b=0$ fm Pb+Pb collisions ({\bf bottom}) with different acceptance applied: full rapidity region ({\bf left}), central rapidity region $|y| <1$ ({\bf center}) and NA61/SHINE acceptance ({\bf right}).}
    \label{fig:omega}
\end{figure*}

The scaled variances $\omega[X]$ are
calculated for distribution of multiplicities of negative $N_-$ and  all charged  $N_{\rm ch}$
particles in different acceptance regions.
Although for mean particle multiplicities $\langle N_{\rm ch} \rangle$ the results of all
mentioned transport  models
are quite in agreement, the values of scaled variance $\omega[X]$ differ significantly.

In Fig. \ref{fig:omega} the calculated values of $\omega[N_-]$ as functions of
$\sqrt{s_{\rm NN}}$ are presented. In the top panel, the results for p+p inelastic
collisions are shown, results for Ar+Sc collisions are presented in the middle panel,
and Pb+Pb in the bottom panel. Both Ar+Sc and Pb+Pb results correspond to the sample
of events with zero impact parameter $b=0$.
The results for full $4\pi$ acceptance, $|y|<1$, and the NA61/SHINE acceptance
are shown in the left, center, and right column, respectively. From results
presented in  Fig. \ref{fig:omega} one concludes:

1. The numerical values of $\omega[N_-]$ appear to be rather different in different transport models.

2. Most model calculations show a monotonous increase of $\omega[N_-]$ with
collision energy.

3. The results for $4\pi$ fluctuations (left column) and those for $|y|<1$ (central column)
show the behavior of $\omega[N_-]$ which is qualitatively different from Eq.\ref{eq:omega-acc},
i.e., for all transport models
the values of $\omega_{\rm acc}[N_-]$ for $|y|<1$ are {\it larger} than $\omega[N_-]$
for full $4\pi$ acceptance.

4. The $4\pi$ values of $\omega[N_-]$ (left column) in Pb+Pb collisions with $b=0$ are
{\it smaller} than the corresponding values in p+p inelastic reactions. This behavior is qualitatively
different from Eq.\ref{part} which says that $\omega[N_-]$ in A+A collisions
should be larger or equal to $\omega_{\rm NN}[N_-]$ in NN collisions.

\subsection{Strongly intensive measures $\Delta$ and $\Sigma$}

\begin{figure*}[t]
\centering
     \includegraphics[width=.7\textwidth]{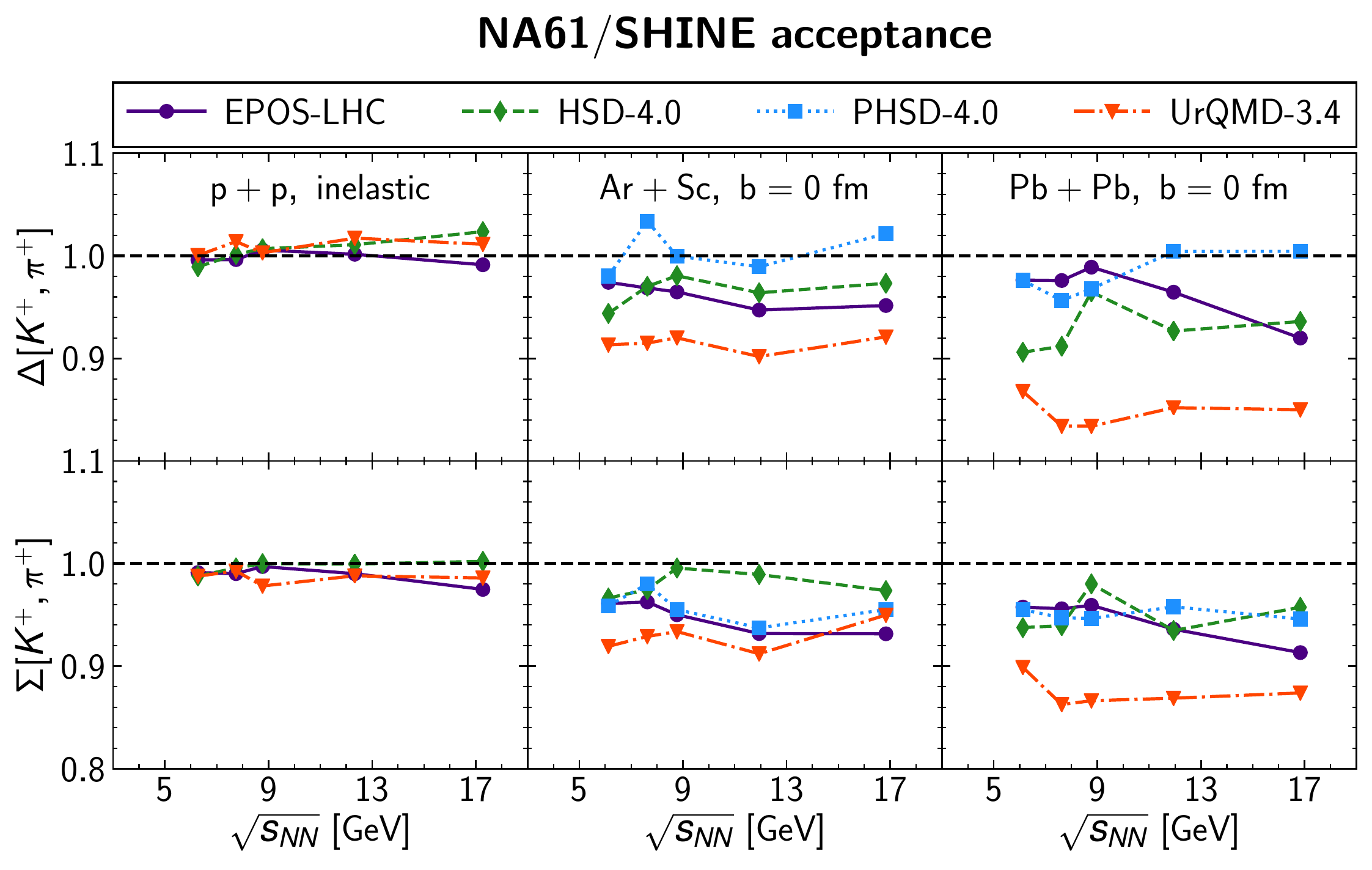}
    \caption{Strongly intensive measures $\Delta[K^+,\pi^+]$ and $\Sigma[K^+,\pi^+]$ in inelastic p+p collisions ({\bf left}), $b=0$ fm Ar+Sc collisions ({\bf middle}),  and $b=0$ fm Pb+Pb collisions ({\bf right}) with NA61/SHINE experimental acceptance applied. Comparison of EPOS (purple circles), HSD (green diamonds),  PHSD (blue squares) and UrQMD (red triangles).}
    \label{fig:sigma-delta}
\end{figure*}

The strongly intensive measures $\Delta$ and $\Sigma$ are expected
to be independent of system volume fluctuations and thus these quantities do not require complicated centrality selection procedure.
The $\Delta[K^+,\pi^+]$ as well as  $\Sigma[K^+,\pi^+]$ were calculated using studied transport models. Their values
are obtained
with NA61/SHINE acceptance applied.
In Fig. \ref{fig:sigma-delta} the calculated values of $\Delta$ and $\Sigma$ are presented. For p+p collisions all models produce values that are tightly located near value of $1$. For Ar+Sc and Pb+Pb collisions
the models do not show so good agreement between each other, as for p+p. With a growth of the system size
the differences between studied models start to increase, and a non-trivial behavior for UrQMD, HSD and PHSD results emerges.

\subsection{Scaled variance in different rapidity regions}

\begin{figure*}[t]
\centering
    \includegraphics[width=.45\textwidth]{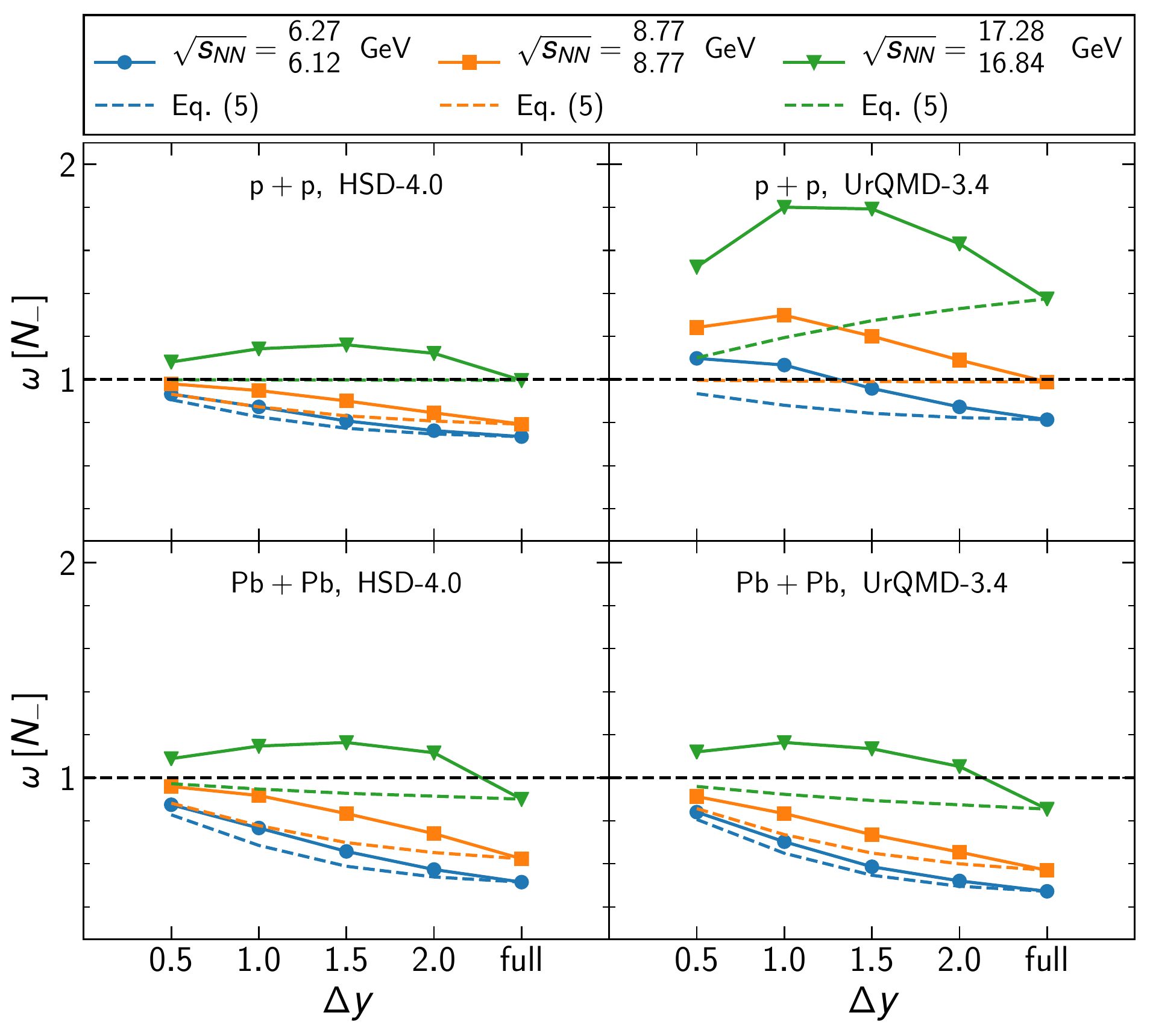}
    \includegraphics[width=.45\textwidth]{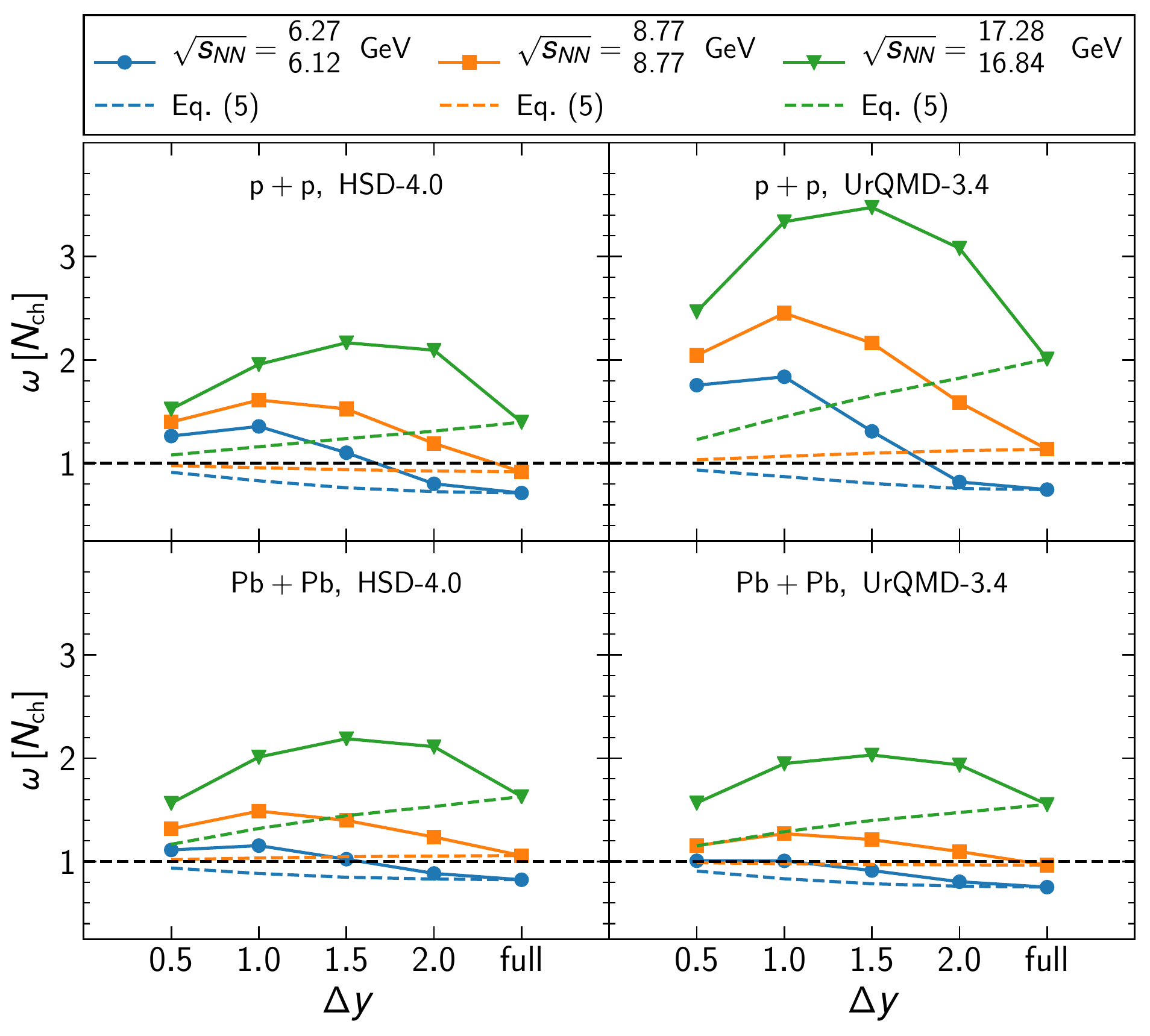}
    \caption{$\omega[N_-]$ (left panel) and $\omega\left[ N_{\rm ch}\right]$
    (right panel)  in inelastic p+p (top) and $b=0$ fm Pb+Pb (bottom) collisions calculated in different rapidity regions within HSD and UrQMD  models
    at different energies
    $\sqrt{s_{NN}}$ (upper numbers correspond to p+p collisions and lower numbers -- to Pb+Pb). The predictions of Eq.\ref{eq:omega-acc} are schematically  presented as dashed lines.}
    	\label{fig:omega_rap_scan}
\end{figure*}

In order to study the acceptance dependence of particle number fluctuations,
the scaled variances $\omega[N_-]$ and  $\omega[N_{\rm ch}]$ have been calculated in
central Pb+Pb and inelastic p+p
reactions within the HSD and UrQMD
models for different rapidity
regions $|y|<\Delta y$ in the center of mass system of colliding protons.
In Fig. \ref{fig:omega_rap_scan}
the calculated scaled variances
$\omega[N_-]$ and $\omega[N_{\rm ch}]$ versus $\Delta y$ are
presented by solid lines. Dashed lines show schematically the behavior which follows from Eq.\ref{eq:omega-acc}.
According to Eq.\ref{eq:omega-acc} the both $\omega[N_-]$ and
$\omega[N_{\rm ch}]$ should be  monotonous functions of $\Delta y$ and should  go to $1$
at $\Delta y\rightarrow 0$.
However, Fig. \ref{fig:omega_rap_scan} shows that $\omega[N_-]$ and
$\omega\left[N_{\rm ch}\right]$ calculated in the HSD and UrQMD models
have a non-monotonous dependence on $\Delta y$.
Therefore,
Eq.\ref{eq:omega-acc}
appears to be in a qualitative  contradiction
with the transport model results.

Note that the non-monotonous behavior is essentially stronger for $\omega[N_{\rm ch}]$.
This is due to decay contributions of heavy meson resonances into $\pi^+\pi^-$ pairs.
Besides, in p+p inelastic collisions UrQMD shows much stronger fluctuations than HSD near the maximum of $\omega[N_{\rm ch}]$.
This can be also explained by a contribution from the resonance decays:
in HSD the particle production at the SPS energies is dominated by string
fragmentation, while in UrQMD a significant part of particles is produced
by formation and decay of resonances.

\subsection{System size dependence and comparison with experimental data}
\label{sec:size_dep}

\begin{figure*}[t]
\includegraphics[width=.7\textwidth]{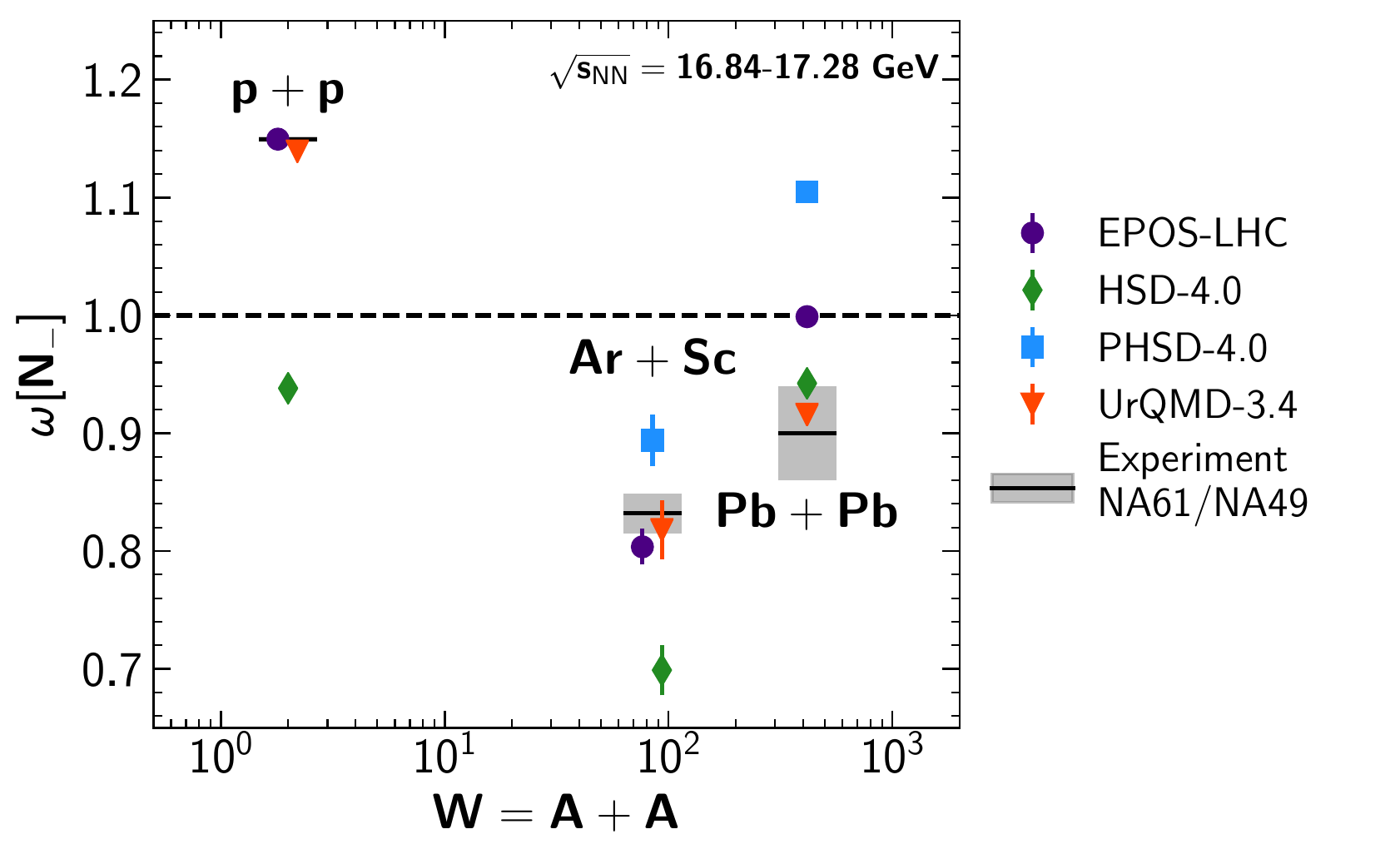}
\caption{Scaled variance of multiplicity distribution of negatively
 charged particles $N_-$ as function of system size (mass number W=A+A of colliding system). Comparison of model results with experimental data. For model
 simulations the experimental acceptance is applied (see text for details).}
\label{fig:omega-size}
\end{figure*}

In Fig. \ref{fig:omega-size} the dependence of $\omega[N_{\rm -}]$ on
the system size  is presented.
The experimental results from \cite{Seryakov:2017sss, Alt:2007jq} are shown for p+p, Ar+Sc,
and Pb+Pb collisions\footnote{Please note, that NA49 acceptance for Pb+Pb point \cite{Alt:2007jq}
was slightly smaller than the acceptance currently used by NA61/SHINE.
In both cases $0<y_\pi<y_{\rm beam}$ rapidity (where pion mass was assumed in calculation of rapidity) range was used but in the NA49 analysis
additional track cuts were used which reduced the mean
multiplicity of negatively charged hadrons by $9\%$ when compared to basic $0<y_\pi<y_{\rm beam}$ cut.}.

The transport models results for p+p collisions presented  in Fig. \ref{fig:omega-size}
correspond to the sample of all inelastic events.
To fulfill experimental centrality selection
conditions in our calculations of Ar+Sc collisions we use
the acceptance maps of the Projectile Spectator
Detector (PSD) \cite{PSD-maps}. For every event in the minimum bias sample
the energy of all particles accepted by the
PSD is calculated and then the distribution of events
over the forward energy is used to select 0.2\% of most central (violent) collisions.
In this centrality class, all transport models produce the values of scaled variance
$\omega[N_-]$ smaller than unity. These results are close to the data
and are in a contradiction with wounded nucleon model given by Eq.\ref{part}.
The $b=0$ fm Pb+Pb collisions are calculated in the transport models
and compared with the 1\% of most central
events. As has been stated in Ref.~\cite{Konchakovski:2007ah, Lungwitz:2007jq}
the HSD and UrQMD results in Pb+Pb collisions with zero impact parameter
can be  a good approximation for the 1\% most central events.

A difference of the multiplicity fluctuations in the central collisions of heavy nuclei
and inelastic p+p collisions is clearly seen in the SPS energy region. For most central
nucleus-nucleus collisions one expects that the system of final particles possesses the properties
of statistical systems. In this case, the global charge conservation strongly suppresses the final 
multiplicities of charged hadrons \cite{Begun:2004gs}. The suppression takes place for large systems (even in the thermodynamic limit)
and survives partially in  any finite parts of the whole phase space. This finally leads to $\omega[N_-]< 1$,
i.e., the particle number distribution for $N_-$ is narrower than the Poisson distribution. To see such an effect,
one needs to make a very rigid centrality selection. Otherwise, the enhancement of $\omega[N_-]$ because of
'volume fluctuations' overcomes the charge conservation
suppression and would lead to $\omega[N_-]>1$. For p+p collisions the situation is different. Firstly, the created system is small,
and for small systems the role of global charge conservation is surprisingly not so strong as for large systems \cite{Begun:2004gs}. However, 
even more important is the fact that the fluctuation  data for all inelastic p+p collisions are combined in the one data set,
i.e., the centrality selection was never done for the multiplicity fluctuations. To clarify the basic physical effects
we propose to present the existing  p+p data within several narrow sets on the `centrality'. To define  the centrality
samples in p+p collisions further checks will be probably required.

\section{Summary}
\label{sec:summary}

The scaled variances of particle number distribution for $N_-$ and $N_{\rm ch}$,
and strongly intensive
measures for fluctuations of $\pi^+$ and $K^+$ numbers are calculated within
several transport models, EPOS, PHSD, and UrQMD, in p+p, Ar+Sc, and Pb+Pb collisions
at the SPS energy region $\sqrt{s_{NN}}=5.1-17.3$~GeV. The numerical values of scaled variances
$\omega[N_-]$ and $\omega[N_{\rm ch}]$ obtained in different models appear to be rather different. On the other hand,
all of the studied transport models do not show any
signs of transition to deconfined phase in fluctuation measures.
Although in the PHSD model the partonic degrees of freedom are present,
and the horn in $K^+/\pi^+$ ratio is reproduced \cite{Palmese:2016rtq},
still there is no non-monotonic behavior of particle number fluctuations as functions of
the beam energy.

Several unexpected properties are observed within the transport model
simulations for scaled variances as functions
of the system size and applied acceptance.
All transport models appear
to be in qualitative contradiction with the often
used formula (\ref{eq:omega-acc}) implementing the acceptance effects.
We observe that, in contrast to Eq.\ref{eq:omega-acc},
$\omega[N_{\rm ch}]$ depends non-monotonously on the size of the acceptance
interval in the central rapidity region.
We also find that in  most central Ar+Sc and Pb+Pb collisions
the scaled variance $\omega[N_-]$ is smaller than
its value in inelastic p+p collisions at the same collision energy $\sqrt{s_{NN}}\cong 17$~GeV.
This result is in a qualitative  contradiction with the model of independent sources (\ref{part}).

Transport models, with the experimental centrality selection and
acceptance of the NA61/SHINE detectors are applied to calculate $\omega[N_-]$
in 0.2\%  most central Ar+Sc collisions. Transport model results give the values
of $\omega[N_-]$ essentially smaller than 1, in agreement with the data.
Transport models give also a satisfactory description for $\omega[N_-]$ of the NA61/SHINE
data in p+p inelastic reactions and for the NA49 data in 1\%  central Pb+Pb collisions.

Obtained results support
a conclusion that a contribution of the quark-gluon degrees of freedom is hardly seen in
the present data on fluctuations at the SPS energies. On the other hand, the NA61/SHINE data
for $\omega[N_-]$ show rather different behavior in p+p inelastic reactions and most central collisions
of heavy ions. It looks rather interesting to investigate a transition region between these
two regimes. We plan to apply transport models to study light nucleus-nucleus collisions at the SPS energies
in our future studies.

\section{Acknowledgements}
\label{sec:acknowledgements}

We are thankful to Larissa Bravina, Marek Ga\`zdzicki,
Tanguy Pierog, Andrey Seryakov, Jan Steinheimer, Horst St\"ocker, and Volodymyr Vovchenko for fruitful discussions.
A.M. acknowledges the support by HGS-HIRe for FAIR, and
Norwegian Centre for International Cooperation in Education, Grant No. CPEA-LT-2016/10094.
The work of K.G. was partially supported by the National Science Centre, Poland grant 2015/18/M/ST2/00125.
The work of M.I.G. is supported by the Program of Fundamental Research of the Department of Physics and Astronomy of National Academy of Sciences of Ukraine.
The work has been performed in the framework of COST Action CA15213 THOR.
The computational resources have been provided by the LOEWE-CSC.


\begin{thebibliography}{99}

\bibitem{Heiselberg:2000fk}
  H.~Heiselberg,
  Phys.\ Rept.\  {\bf 351}, 161 (2001)
  doi:10.1016/S0370-1573(00)00140-X
  [nucl-th/0003046].


\bibitem{Jeon:2003gk}
  S.~Jeon and V.~Koch,\label{eq:omega-wound-nucl}
  In *Hwa, R.C. (ed.) et al.: Quark gluon plasma* 430-490
  [hep-ph/0304012].


\bibitem{Asakawa:2015ybt}
  M.~Asakawa and M.~Kitazawa,
  Prog.\ Part.\ Nucl.\ Phys.\  {\bf 90}, 299 (2016)
  doi:10.1016/j.ppnp.2016.04.002
  [arXiv:1512.05038 [nucl-th]].

\bibitem{Werner:2005jf}
  K.~Werner, F.~M.~Liu and T.~Pierog,
  Phys.\ Rev.\ C {\bf 74}, 044902 (2006)
  doi:10.1103/PhysRevC.74.044902
  [hep-ph/0506232].


\bibitem{Werner:2010aa}
  K.~Werner, I.~Karpenko, T.~Pierog, M.~Bleicher and K.~Mikhailov,
  Phys.\ Rev.\ C {\bf 82}, 044904 (2010)
  doi:10.1103/PhysRevC.82.044904
  [arXiv:1004.0805 [nucl-th]].


\bibitem{Cassing:2008sv}
  W.~Cassing and E.~L.~Bratkovskaya,
  Phys.\ Rev.\ C {\bf 78}, 034919 (2008)
  doi:10.1103/PhysRevC.78.034919
  [arXiv:0808.0022 [hep-ph]].


\bibitem{Cassing:2009vt}
  W.~Cassing and E.~L.~Bratkovskaya,
  Nucl.\ Phys.\ A {\bf 831}, 215 (2009)
  doi:10.1016/j.nuclphysa.2009.09.007
  [arXiv:0907.5331 [nucl-th]].


\bibitem{Bass:1998ca}
  S.~A.~Bass {\it et al.},
  Prog.\ Part.\ Nucl.\ Phys.\  {\bf 41}, 255 (1998)
  [Prog.\ Part.\ Nucl.\ Phys.\  {\bf 41}, 225 (1998)]
  doi:10.1016/S0146-6410(98)00058-1
  [nucl-th/9803035].


\bibitem{Bleicher:1999xi}
  M.~Bleicher {\it et al.},
  J.\ Phys.\ G {\bf 25}, 1859 (1999)
  doi:10.1088/0954-3899/25/9/308
  [hep-ph/9909407].


\bibitem{Konchakovski:2005hq}
  V.~P.~Konchakovski, S.~Haussler, M.~I.~Gorenstein, E.~L.~Bratkovskaya, M.~Bleicher and H.~Stoecker,
  Phys.\ Rev.\ C {\bf 73}, 034902 (2006)
  doi:10.1103/PhysRevC.73.034902
  [nucl-th/0511083].


\bibitem{Konchakovski:2006aq}
  V.~P.~Konchakovski, M.~I.~Gorenstein, E.~L.~Bratkovskaya and H.~Stoecker,
  Phys.\ Rev.\ C {\bf 74}, 064911 (2006)
  doi:10.1103/PhysRevC.74.064911
  [nucl-th/0606047].


\bibitem{Gorenstein:2008et}
  M.~I.~Gorenstein, M.~Hauer, V.~P.~Konchakovski and E.~L.~Bratkovskaya,
  Phys.\ Rev.\ C {\bf 79}, 024907 (2009)
  doi:10.1103/PhysRevC.79.024907
  [arXiv:0811.3089 [nucl-th]].


\bibitem{Konchakovski:2010fh}
  V.~P.~Konchakovski, M.~I.~Gorenstein, E.~L.~Bratkovskaya and W.~Greiner,
  J.\ Phys.\ G {\bf 37}, 073101 (2010)
  doi:10.1088/0954-3899/37/7/073101
  [arXiv:1001.3085 [nucl-th]].


\bibitem{Begun:2012wq}
  V.~V.~Begun, V.~P.~Konchakovski, M.~I.~Gorenstein and E.~Bratkovskaya,
  J.\ Phys.\ G {\bf 40}, 045109 (2013)
  doi:10.1088/0954-3899/40/4/045109
  [arXiv:1205.6809 [nucl-th]].


\bibitem{Vovchenko:2014ssa}
  V.~Y.~Vovchenko, D.~V.~Anchishkin and M.~I.~Gorenstein,
  Phys.\ Rev.\ C {\bf 90}, no. 2, 024916 (2014)
  doi:10.1103/PhysRevC.90.024916
  [arXiv:1407.0629 [nucl-th]].


\bibitem{Steinheimer:2016cir}
  J.~Steinheimer, V.~Vovchenko, J.~Aichelin, M.~Bleicher and H.~St\"ocker,
  arXiv:1608.03737 [nucl-th].


\bibitem{Seryakov:2017sss}
  A.~Seryakov [NA61/SHINE Collaboration],
  Acta Phys. Polon. B Proc. Suppl. {\bf 10}, 723 (2017)
  [arXiv:1704.00751 [hep-ex]].


\bibitem{Antoniou:2006mh}
  N.~Antoniou {\it et al.} [NA49-future Collaboration],
  CERN-SPSC-2006-034, CERN-SPSC-P-330.


\bibitem{Abgrall:2014xwa}
  N.~Abgrall {\it et al.} [NA61 Collaboration],
  JINST {\bf 9}, P06005 (2014)
  doi:10.1088/1748-0221/9/06/P06005
  [arXiv:1401.4699 [physics.ins-det]].

\bibitem{Seryakov-private}
	A.~Seryakov, (2017), private communication

\bibitem{Gorenstein:2011vq}
  M.~I.~Gorenstein and M.~Gazdzicki,
  Phys.\ Rev.\ C {\bf 84}, 014904 (2011)
  doi:10.1103/PhysRevC.84.014904
  [arXiv:1101.4865 [nucl-th]].


\bibitem{Begun:2004zb}
  V.~V.~Begun, M.~I.~Gorenstein and O.~S.~Zozulya,
  Phys.\ Rev.\ C {\bf 72}, 014902 (2005)
  doi:10.1103/PhysRevC.72.014902
  [nucl-th/0411003].


\bibitem{Bialas:1976ed}
  A.~Bialas, M.~Bleszynski and W.~Czyz,
  Nucl.\ Phys.\ B {\bf 111}, 461 (1976)
  doi:10.1016/0550-3213(76)90329-1


\bibitem{Drescher:2000ha}
  H.~J.~Drescher, M.~Hladik, S.~Ostapchenko, T.~Pierog and K.~Werner,
  Phys.\ Rept.\  {\bf 350}, 93 (2001)
  doi:10.1016/S0370-1573(00)00122-8
  [hep-ph/0007198].


\bibitem{Bratkovskaya:2011wp}
  E.~L.~Bratkovskaya, W.~Cassing, V.~P.~Konchakovski and O.~Linnyk,
  Nucl.\ Phys.\ A {\bf 856}, 162 (2011)
  doi:10.1016/j.nuclphysa.2011.03.003
  [arXiv:1101.5793 [nucl-th]].


\bibitem{Cassing:2008nn}
  W.~Cassing,
  Eur.\ Phys.\ J.\ ST {\bf 168}, 3 (2009)
  doi:10.1140/epjst/e2009-00959-x
  [arXiv:0808.0715 [nucl-th]].


\bibitem{Kadanoff1}
  L. P. Kadanoff and G. Baym,  {\it Quantum Statistical Mechanics},
  Benjamin, New York, 1962.


\bibitem{Cassing:1999es}
  W.~Cassing and E.~L.~Bratkovskaya,
  Phys.\ Rept.\  {\bf 308}, 65 (1999)
  doi:10.1016/S0370-1573(98)00028-3


\bibitem{Cassing:2000bj}
 W.~Cassing, E.~L.~Bratkovskaya and S.~Juchem,
 Nucl.\ Phys.\ A {\bf 674}, 249 (2000)
 doi:10.1016/S0375-9474(00)00163-9
 [nucl-th/0001024].

\bibitem{Berrehrah:2016vzw}
  H.~Berrehrah, E.~Bratkovskaya, T.~Steinert and W.~Cassing,
  Int.\ J.\ Mod.\ Phys.\ E {\bf 25}, no. 07, 1642003 (2016)
  doi:10.1142/S0218301316420039
  [arXiv:1605.02371 [hep-ph]].

\bibitem{Ozvenchuk:2012kh}
  V.~Ozvenchuk, O.~Linnyk, M.~I.~Gorenstein, E.~L.~Bratkovskaya and W.~Cassing,
  Phys.\ Rev.\ C {\bf 87}, no. 6, 064903 (2013)
  doi:10.1103/PhysRevC.87.064903
  [arXiv:1212.5393 [hep-ph]].


\bibitem{Andersson:1992iq}
  B.~Andersson, G.~Gustafson and H.~Pi,
  Z.\ Phys.\ C {\bf 57}, 485 (1993)
  doi:10.1007/BF01474343


\bibitem{Konchakovski:2014fya}
  V.~P.~Konchakovski, W.~Cassing and V.~D.~Toneev,
  J.\ Phys.\ G {\bf 42}, no. 5, 055106 (2015)
  doi:10.1088/0954-3899/42/5/055106
  [arXiv:1411.5534 [nucl-th]].


\bibitem{Konchakovski:2011qa}
  V.~P.~Konchakovski, E.~L.~Bratkovskaya, W.~Cassing, V.~D.~Toneev and V.~Voronyuk,
  Phys.\ Rev.\ C {\bf 85}, 011902 (2012)
  doi:10.1103/PhysRevC.85.011902
  [arXiv:1109.3039 [nucl-th]].

\bibitem{Linnyk:2015rco}
  O.~Linnyk, E.~L.~Bratkovskaya and W.~Cassing,
  Prog.\ Part.\ Nucl.\ Phys.\  {\bf 87}, 50 (2016)
  doi:10.1016/j.ppnp.2015.12.003
  [arXiv:1512.08126 [nucl-th]].

\bibitem{Palmese:2016rtq}
  A.~Palmese, W.~Cassing, E.~Seifert, T.~Steinert, P.~Moreau and E.~L.~Bratkovskaya,
  Phys.\ Rev.\ C {\bf 94}, no. 4, 044912 (2016)
  doi:10.1103/PhysRevC.94.044912
  [arXiv:1607.04073 [nucl-th]].

\bibitem{NA61-acceptance}
	Acceptance maps used in the paper: ``Multiplicity and transverse momentum fluctuations in proton-proton interactions,'' \href{https://edms.cern.ch/document/1549298/1}{https://edms.cern.ch/document/1549298/1}

\bibitem{Alt:2007jq}
  C.~Alt {\it et al.} [NA49 Collaboration],
  Phys.\ Rev.\ C {\bf 78}, 034914 (2008)
  doi:10.1103/PhysRevC.78.034914
  [arXiv:0712.3216 [nucl-ex]].

\bibitem{PSD-maps}
A.~Seryakov, \href{https://edms.cern.ch/document/1867336/1}{https://edms.cern.ch/document/1867336/1}

\bibitem{Lungwitz:2007jq}
  B.~Lungwitz {\it et al.} [NA49 Collaboration],
  Phys.\ Rev.\ C {\bf 78}, 034914 (2008)
  doi:10.1103/PhysRevC.78.034914
  [arXiv:0712.3216 [nucl-ex]].

\bibitem{Konchakovski:2007ah}
  V.~P.~Konchakovski, B.~Lungwitz, M.~I.~Gorenstein and E.~L.~Bratkovskaya,
  Phys.\ Rev.\ C {\bf 78}, 024906 (2008)
  doi:10.1103/PhysRevC.78.024906
  [arXiv:0712.2044 [nucl-th]].
  
\bibitem{Begun:2004gs} 
  V.~V.~Begun, M.~Gazdzicki, M.~I.~Gorenstein and O.~S.~Zozulya,
  Phys.\ Rev.\ C {\bf 70}, 034901 (2004)
  doi:10.1103/PhysRevC.70.034901
  [nucl-th/0404056].
\end{thebibliography}
\end{document}